\documentclass[12pt]{iopart}
\usepackage{amssymb}
\usepackage{graphicx,subfig}

\begin{document}
\title{Model- and calibration-independent test of cosmic acceleration}
\author{Marina Seikel$^1$ and Dominik J Schwarz$^2$}
\address{Fakult\"at f\"ur Physik, Universit\"at Bielefeld, Postfach
  100131, 33501 Bielefeld, Germany}
\ead{\mailto{$^1$mseikel@physik.uni-bielefeld.de},
  \mailto{$^2$dschwarz@physik.uni-bielefeld.de}} 

\begin{abstract}
We present a calibration-independent test of the accelerated
expansion of the universe using supernova type Ia data. The test is
also model-independent in the sense that no assumptions about the
content of the universe or about the parameterization of the
deceleration parameter are made and that it does not assume any
dynamical equations of motion. Yet, the test assumes the 
universe and the distribution of supernovae to be
statistically homogeneous and isotropic. 
A significant reduction of systematic effects, as compared to our
previous, calibration-dependent test, is achieved.
Accelerated expansion is detected at significant level ($4.3 \sigma$
in the 2007 Gold sample, $7.2 \sigma$ in the 2008 Union sample) if the 
universe is spatially flat. This result depends,
however, crucially on supernovae with a redshift smaller than 0.1, for
which the assumption of statistical isotropy and homogeneity is less
well established.
\end{abstract}


\section{Introduction}
The concordance model of cosmology is very successful in describing
observations. It states that the universe consists of baryonic and
cold dark matter as well as a cosmological constant, where the
baryonic matter makes up only 5\% of the content of the universe. The
cosmological constant contributes 72\% and thus causes an accelerated
expansion at the present epoch. A large variety of other cosmological
models have been proposed that come to similar results. Especially as
soon as homogeneity and isotropy are assumed, each model states
accelerated expansion of the universe when confronted with
observational data.

If we are only interested in the question, whether the universe really
expands accelerated, but not in the specific content of the universe,
proposing and testing a certain cosmological model is not the
appropriate approach. This question rather needs to be answered as
model-independent as possible, i.e. without making any assumptions
about the matter and energy content of the universe. The so-called
kinematical approach does so by using special parameterizations of the
deceleration parameter $q(z)$ \cite{Turner,Riess04,Elgaroy},
the scale parameter $a(t)$ \cite{Wang}, the Hubble rate $H(z)$
\cite{John}, the dimensionless coordinate distance $y(z)$
\cite{Daly03, Daly08} or different distance scales \cite{Cattoen}. 
Other published methods are to expand $q$ into
principle components \cite{Shapiro} or to expand the jerk parameter
$j$ into a series of orthonormal functions \cite{Rapetti}.

It is, however, even possible to avoid these kinds of
parameterizations. In \cite{Seikel} we presented a model-independent
test to quantify the evidence for accelerated expansion with the only
assumption that the universe is homogeneous and isotropic. While
versions of this test have already been proposed \cite{Visser} and
applied \cite{Santos,Gong} by other groups, we additionally considered
calibration effects, quantified the evidence for accelerated expansion
and studied systematic effects. 

The crucial assumption in our test is the Copernican principle. 
It states that we are typical observers in the universe. Together with
the observed isotropy of the cosmic microwave background, 
the Copernican principle implies that the universe is also  
homogeneous, the statement of the cosmological principle. 

Although it seems to be consequent to adopt the cosmological principle
for the analysis of supernovae, a critical reflection on it is in
order. Most probes of the statistical isotropy use objects at rather
high redshift, while nearby probes show significantly less evidence 
for statistical isotropy. For supernovae, this issue has been investigated 
recently in \cite{Haugboelle,SW}. Statistically significant violation
of isotropy was found for supernovae at redshift $z < 0.2$ \cite{SW}:  
The fluctuation of the Hubble rate $\Delta H/H$ on opposite
hemispheres on the sky is about $5\%$. This corresponds to an anisotropy of
distance moduli of $0.1$ mag, which has to be compared to the effect
of acceleration, which is about $0.2$ mag. 

The statistical homogeneity of the universe is less well established 
than the statistical isotropy at high redshifts. Several observations
indicate that the scale of statistical homogeneity is of the order of 
$100$ Mpc (e.g.~from luminous red galaxies \cite{Hogg}, but see also
\cite{Labini}). For supernovae it has been pointed out already many years
ago that dropping the assumption of homogeneity, but keeping that of
isotropy around one point, allows for excellent fits of the Hubble diagram 
without invoking dark energy \cite{Celerier,CelerierRev}. This does
not come as a surprise, because spherically symmetric
dust models (also called Lemaitre-Tolman-Bondi models)
have two arbitrary free functions and if only supernovae are considered
a perfect fit can be obtained. Thus instead of interpreting the supernovae
Hubble diagrams as evidence for dark energy, one could question the 
Copernican principle. In our test, we assume statistical 
homogeneity and isotropy. We somewhat relax that assumption in a second
step by excluding nearby supernovae from the test.
We will demonstrate below, that the nearby supernovae at $z<0.1$ are
crucial in the detection of cosmic acceleration. 

In the following section, we will
shortly summarize the basic concept and results of our test.
In the present work, we slightly modify the test in order to avoid
systematics due to calibration. The reason for those systematics is
also explained in the next section. In section \ref{calibration-free}
the modified test is presented and applied to different data sets,
including the recently published Union set \cite{Kowalski}, assuming a
spatially flat universe. The cases of open and closed universes will
be considered in section \ref{open-closed}.

\section{Calibration-dependent test}
We consider the distance modulus 
\begin{equation}\label{mueq}
\mu=m-M = 5\log d_{\mbox{\scriptsize L}} +25 
\end{equation}
of supernovae type Ia (SN Ia), where the luminosity distance
$d_{\mbox{\scriptsize L}}$ is given in units of Mpc. $m$ and $M$ are
the apparent and absolute magnitudes, respectively. 
If $\mu_i$ is the distance modulus of the $i$-th SN with redshift
$z_i$, one can define a new quantity
\begin{equation}\label{deltamui}
\Delta\mu_i = \mu_i - \mu_{q=0}(z_i) = \mu_i - 5\log\left[
  \frac{1}{H_0}(1+z_i)\ln(1+z_i) \right] -25 \,,
\end{equation}
where $\mu_{q=0}(z)$ is the distance modulus of a universe that
neither accelerates nor decelerates, i.e. with deceleration $q(z)=0$. 

The weighted average of the $\Delta\mu_i$ is given by
\begin{equation}
\Delta\mu =
\frac{\sum_{i=1}^Ng_i\Delta\mu_i}{\sum_{i=1}^Ng_i} \;,
\end{equation}
where $g_i=1/\sigma_i^2$. The $\sigma_i$ include measurement errors
and errors due to peculiar velocities.
The standard deviation of this average is calculated by
\begin{equation}
\sigma =
\left[\frac{\sum_{i=1}^Ng_i\left(\Delta\mu_i -
    \Delta\mu\right)^2}{(N-1)\sum_{i=1}^Ng_i}\right]^{\frac{1}{2}}
\;,
\end{equation}
with $N$ being the number of SNe that is averaged over.

Our null hypothesis is that the universe never expanded accelerated
which implies 
\begin{equation}\label{hypo}
  \Delta\mu \le 0\;. 
\end{equation}
Note that this holds independently of
the content of the universe and does not depend on the validity of
Einstein's equations. Thus if the observed value of
$\Delta\mu$ is significantly larger than zero, the null hypothesis can
be rejected. In that case, one can state that there must have been a
phase of acceleration. However, this does not exclude a phase of
deceleration. From \eref{mueq} and \eref{deltamui} it follows that we
need $M$ and $H_0$ in order to test the null hypothesis
\eref{hypo}. There always exist values of $M$ and $H_0$ such that
\eref{hypo} is satisfied. Thus, these values have to be fixed by an
independent calibration measurement. This is in contrast to
model-dependent tests, where a combination of $M$ and $H_0$ is used as
a fitting parameter.

In \cite{Seikel}, we considered two SN Ia data sets (the Gold sample
\cite{Riess07} and the ESSENCE set \cite{WoodVasey}), two different
light-curve fitters (MLCS2k2 \cite{Jha} and SALT \cite{Guy}) and two
different calibrations (the calibration presented by Riess et
al.~\cite{Riess05} and that given by Sandage et
al.~\cite{Sandage}, which in the following will be referred to
as Riess calibration and Sandage calibration, respectively). We
calculated the averaged value
$\Delta\mu$ over all SNe of a data set with a redshift $z\ge0.2$ and
divided the result by its standard deviation $\sigma$, thus obtaining
the evidence for accelerated expansion. Assuming a flat universe, the
ESSENCE set using the MLCS2k2 fitter and the Sandage calibration gave
the weakest evidence, namely $5.2\sigma$. In the other cases the
evidence is much larger and goes up to 17$\sigma$ for ESSENCE (SALT)
in the Riess calibration. Thus, we observe enormous systematic effects
for the different data sets, fitting methods and calibrations.

Consequently, we have to make an attempt to reduce these large
systematics. It turns out that they are largely due to systematics in
the calibration. This can be understood by the following
considerations.

The absolute magnitude $M$ and the Hubble constant $H_0$
cannot be determined independently by only considering SNe. Thus, the
absolute magnitude of the SNe Ia has to be calibrated by measurements
of the distance moduli of cepheids in the host galaxies. Then SNe can be used
to determine $H_0$. As there is still some controversy between
different groups about the correct calibration, we considered two very
discrepant calibrations for our test of accelerated expansion. 
This test, however, does not depend on $M$ and $H_0$
independently, but on the quantity $\mathcal{M}=M-5\log(H_0)+25$, which
can be seen when we rewrite the null hypothesis $\Delta\mu \le 0$ as
\begin{equation}\label{Mhypo}
m-5\log\left[(1+z)\ln(1+z)\right] \le \mathcal{M}\,.
\end{equation}

The fact that we observe huge systematic errors depending on the
considered calibration seems somewhat strange: 
Assume we have found two different values $M_1$ and $M_2$ for the
absolute magnitude of SNe Ia by cepheid measurements. Using these
results, two values $H_{01}$ and $H_{02}$ for the Hubble constant can
be obtained by observations of nearby SNe. Although $M_1\neq M_2$ and
$H_{01}\neq H_{02}$, the resulting values $\mathcal{M}_1$ and
$\mathcal{M}_2$ are equal by definition, if the same set of low
redshift SNe and
the same analysis is used for the determination of the Hubble constant. 

\begin{center}
\setlength{\unitlength}{1em}
\begin{picture}(15,4)
\put(0,3){$M_1$}
\put(0,0){$M_2$}
\put(1.6,3.4){\vector(2,-1){1.5}}
\put(1.6,0.5){\vector(2,1){1.5}}
\put(4.1,1.6){SNe}
\put(5,2){\circle{3}}
\put(6.9,2.7){\vector(2,1){1.5}}
\put(6.9,1.2){\vector(2,-1){1.5}}
\put(8.7,3){$H_{01}$}
\put(8.7,0){$H_{02}$}
\thicklines
\put(10.9,3.4){\vector(2,-1){1.5}}
\put(10.9,0.5){\vector(2,1){1.5}}
\put(13,1.6){$\mathcal{M}$}
\end{picture}
\end{center}

For our test in \cite{Seikel}, we adopted the values of $M$ and $H_0$
given by Riess et al.~\cite{Riess05} and Sandage et
al.~\cite{Sandage}. As the two groups analysed different SNe and used
different analysis pipelines, they obtained $H_{01}$ and $H_{02}$
which (combined with $M_1$ and $M_2$) did not lead to the same
$\mathcal{M}$, but different values $\mathcal{M}_1$ and $\mathcal{M}_2$.

\begin{center}
\setlength{\unitlength}{1em}
\begin{picture}(15,4)
\put(0,2.8){$M_1$}
\put(0,0.4){$M_2$}
\put(1.6,3.2){\vector(1,0){1.5}}
\put(1.6,0.8){\vector(1,0){1.5}}
\put(4.,2.8){SNe$_1$}
\put(4.,0.4){SNe$_2$}
\put(5,3.2){\oval(3,1.5)}
\put(5,0.8){\oval(3,1.5)}
\put(6.9,3.2){\vector(1,0){1.5}}
\put(6.9,0.8){\vector(1,0){1.5}}
\put(8.7,2.8){$H_{01}$}
\put(8.7,0.4){$H_{02}$}
\thicklines
\put(10.9,3.2){\vector(1,0){1.5}}
\put(10.9,0.8){\vector(1,0){1.5}}
\put(13,2.8){$\mathcal{M}_1$}
\put(13,0.4){$\mathcal{M}_2$}
\end{picture}
\end{center}
Thus, the observed systematics are not due to a different determination
of the absolute magnitude $M$, but are caused by the systematic
errors and the different SN data sets used in the measurement of
$\mathcal{M}$.

While our approach is to test a null hypothesis, model-dependent tests
fit special cosmological models or parameterizations to observational
data. An essential difference between these two approaches is that we
test an inequality, whereas model-dependent tests use an equality,
which requires different analyses. Model-dependent tests are typically
based on the minimization of $\chi^2(p_i)$, where $\mathcal{M}$ is one
of the model parameters $p_i$. Then one can estimate the
likelihood as a function of the parameters, which allows
marginalization over $\mathcal{M}$. As in our test an inequality is
tested, this kind of analysis is not suitable. We cannot use
$\mathcal{M}$ as a free parameter, since the null hypothesis
\eref{Mhypo} is always fulfilled for large enough $\mathcal{M}$. Thus
for the test summarized in this section, $\mathcal{M}$ needs to be
calibrated.

\section{Calibration-independent test}\label{calibration-free}

It is, however, easy to modify our test in such a way that we can
avoid using a certain calibration of $\mathcal{M}$ when testing
the accelerated expansion. We just need to consider relative values of
$\Delta\mu$ instead of absolute values, i.e. we use
$\Delta\mu-\Delta\mu_{\mbox{\scriptsize nearby}}$ rather than
$\Delta\mu$, where $\Delta\mu_{\mbox{\scriptsize nearby}}$ is the
average of $\Delta\mu_i$ using only nearby SNe of a data set. Thus,
the null hypothesis now reads\footnote{Note that this hypothesis does
  {\em not} correspond to the hypothesis of a never acclererating
  universe tested by an observer at $z_{\mbox{\scriptsize nearby}}$.}
\begin{equation}
  \Delta\mu-\Delta\mu_{\mbox{\scriptsize nearby}}\le 0\;.
\end{equation}
The standard deviation $\sigma$ is obtained by adding the standard
deviations of $\Delta\mu$ and $\Delta\mu_{\mbox{\scriptsize nearby}}$
in quadrature. 
In the following, we will assume a spatially flat universe. Open and
closed universes will be considered in the next section.

\begin{figure}
\subfloat[Different cosmological models.]{\label{models}
  \includegraphics*[width=7.5cm]{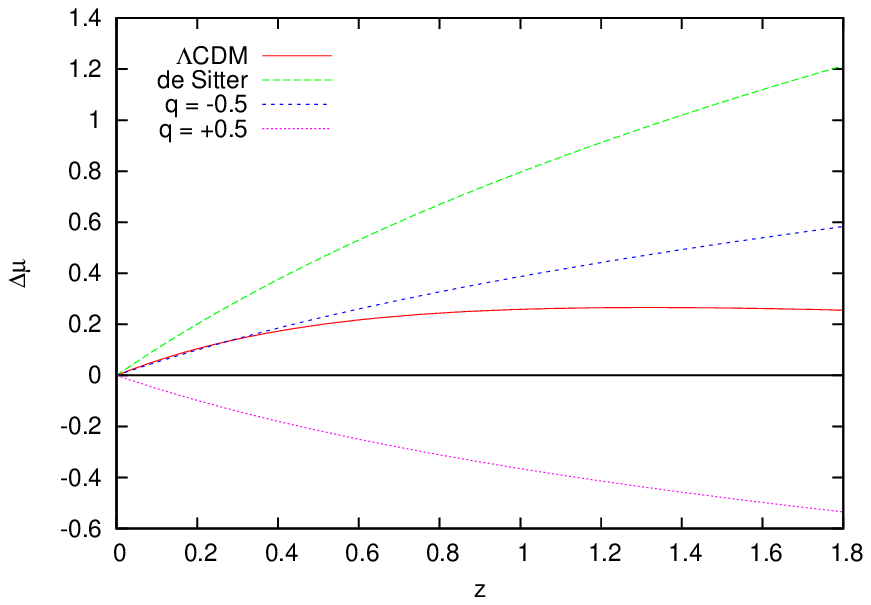}} \hfill
\subfloat[Nearby SNe defined by $0.0\le
  z_{\mbox{\scriptsize nearby}} < 0.2$.]{\label{deltamufiga}
  \includegraphics*[width=7.5cm]{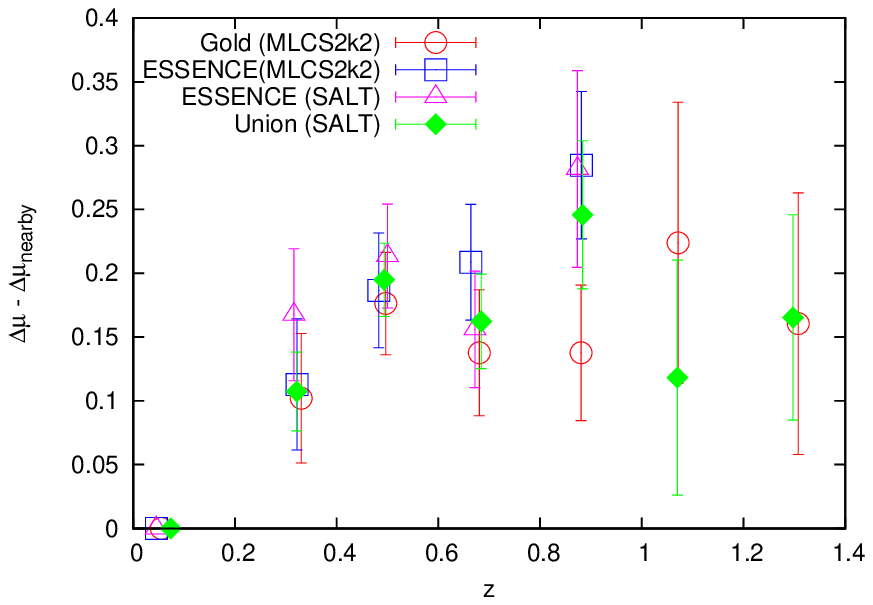}}\\
\subfloat[Nearby SNe defined by $0.0\le
  z_{\mbox{\scriptsize nearby}} < 0.1$.]{\label{deltamufigb}
  \includegraphics*[width=7.5cm]{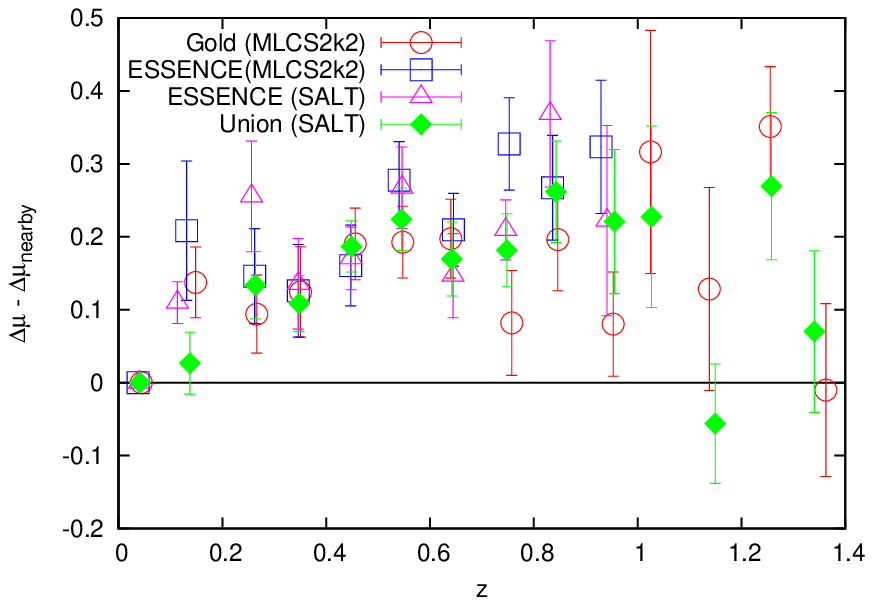}} \hfill
\subfloat[Nearby SNe defined by $0.1\le
  z_{\mbox{\scriptsize nearby}} < 0.2$.]{\label{deltamufigc}
  \includegraphics*[width=7.5cm]{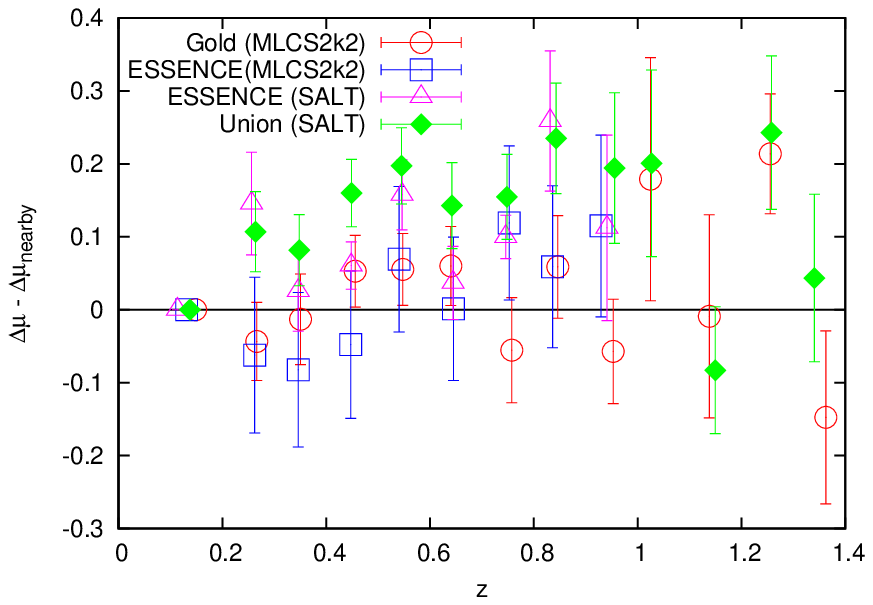}}
\caption{$\Delta\mu$ for different cosmological models (a) and
  $\Delta\mu-\Delta\mu_{\mbox{\scriptsize nearby}}$ for different data
  sets and fitting methods, where nearby SNe are defined as
  those SNe with redshifts fulfilling the given inequalities for
  $z_{\mbox{\scriptsize nearby}}$ (b-d).}
\label{deltamufig}
\end{figure}

In figure \ref{models}, $\Delta\mu(z)$ is plotted for different
spatially flat cosmological models: a $\Lambda$CDM model with
$\Omega_m=0.28$ and $\Omega_\Lambda=0.72$ (WMAP 5yr + BAO + SN best
fit \cite{Komatsu}), a de Sitter universe and
models with constant deceleration ($q=0.5$, which corresponds to the
Einstein-de Sitter model) and constant acceleration
($q=-0.5$). 
Although the transition redshift between acceleration and
deceleration in the $\Lambda$CDM model is about $0.73$, the curve
still slightly increases beyond that redshift and has its maximum at
$z=1.3$. Thus, an increase of $\Delta\mu$ with redshift does not
necessarily correspond to a phase of acceleration at that specific
redshift. Note that for each value of the spatial curvature $\Omega_k$
a different plot is necessary as in general $\Delta\mu_{q=0}(z)$
depends on $\Omega_k$ (see section \ref{open-closed}).

Figure \ref{deltamufiga} shows $\Delta\mu-\Delta\mu_{\mbox{\scriptsize nearby}}$
averaged over redshift bins with width 0.2
for the Gold sample \cite{Riess07} (fitted with MLCS2k2), the
ESSENCE set \cite{WoodVasey} (once fitted with MLCS2k2 and once
with SALT) and the Union set \cite{Kowalski} (fitted with SALT).
As the first bin corresponds to the nearby SNe, its value is per
definition equal to zero. The values in all the other bins are
significantly above zero which indicates accelerated expansion.
It is kind of a natural choice to define the nearby
SNe as all SNe having a redshift smaller than 0.2 since at this point
there is a gap in all presently available data sets. Doing so, there
is a strong evidence for accelerated expansion.

Nevertheless, we also consider the cases where the nearby SNe
are defined as those with $z<0.1$ and those with $0.1\le z<0.2$
(figures \ref{deltamufigb} and \ref{deltamufigc}, respectively).
In both plots a bin width of 0.1 is used. In figure \ref{deltamufigc},
the first bin is skipped and thus the second bin is fixed to zero. One
can already see without any quantitative analysis that the
evidence for acceleration dramatically decreases if SNe with $z<0.1$
are not used for the test. In particular for the two data sets that
are fitted with MLCS2k2, the evidence completely vanishes since
several data points become negative. However, one should not take that
result very serious, as it is based on a very small number of SNe in
the calibrating bin. Most reliable is the Union set with 6 SNe between
redshift 0.1 and 0.2.

A quantitative value for the evidence of acceleration can be obtained
by dividing $\Delta\mu-\Delta\mu_{\mbox{\scriptsize nearby}}$ by its standard
deviation $\sigma$. The results corresponding to the bins plotted in
figure \ref{deltamufig} are given in table \ref{bindeltamutab}.
A better statistics can be achieved when averaging $\Delta\mu$ over
all SNe of a set with a redshift larger than that of  the nearby
ones. The results are listed in table \ref{deltamutab}. Although the
test was only modified in order to avoid
systematics from calibration, also the other systematics are reduced.
In the test presented in \cite{Seikel} the calculated evidences
varied from $11.9\sigma$ to $17.0\sigma$ in the Riess calibration and
from $5.2\sigma$ to $10.4$ in the Sandage calibration. We then only
used the Gold and ESSENCE sets, but not the Union set. For the same
sets, using $z_{\mbox{\scriptsize nearby}}<0.2$, we now obtain values 
that lie much closer to each other, namely between $4.3\sigma$ and
$5.7\sigma$.

\begin{table}
\begin{indented}
\lineup
\caption{\label{bindeltamutab} Evidence for acceleration
  $(\Delta\mu-\Delta\mu_{\mbox{\scriptsize nearby}})/\sigma$ for different data
  sets and fitting methods using SNe in different redshift
  bins. Nearby SNe are defined as 
  those SNe with redshifts fulfilling the given inequalities for
  $z_{\mbox{\scriptsize nearby}}$.}
\item[] 
\begin{tabular}{@{}llllll}
\br
Nearby SNe & Bin  & Gold    & ESSENCE & ESSENCE & Union \\
           &      &(MLCS2k2)&(MLCS2k2)& (SALT)  & (SALT)\\\mr
$0.0\le z_{\mbox{\scriptsize nearby}}<0.2$ & 
 $0.2\le z < 0.4$ & 2.0     & 2.2     & 3.2     & 3.5 \\
&$0.4\le z < 0.6$ & 4.4     & 4.2     & 5.2     & 6.8 \\
&$0.6\le z < 0.8$ & 2.8     & 4.6     & 3.4     & 4.4 \\
&$0.8\le z < 1.0$ & 2.6     & 4.9     & 3.7     & 4.2 \\
&$1.0\le z < 1.2$ & 2.0     &         &         & 1.3 \\
&$1.2\le z < 1.4$ & 1.6     &         &         & 2.1 \\
\mr
$0.0\le z_{\mbox{\scriptsize nearby}}<0.1$ & 
 $0.1\le z < 0.2$ &   2.8   & 2.2     & 3.8     &  0.6 \\
&$0.2\le z < 0.3$ &   1.8   & 2.2     & 3.4     &  2.9 \\
&$0.3\le z < 0.4$ &   2.0   & 2.0     & 2.2     &  2.8 \\
&$0.4\le z < 0.5$ &   3.9   & 2.9     & 4.0     &  5.3 \\
&$0.5\le z < 0.6$ &   3.9   & 5.3     & 4.8     &  5.2 \\
&$0.6\le z < 0.7$ &   3.7   & 4.2     & 2.5     &  3.3 \\
&$0.7\le z < 0.8$ &   1.1   & 5.2     & 5.1     &  3.6 \\
&$0.8\le z < 0.9$ &   2.8   & 3.7     & 3.7     &  3.8 \\
&$0.9\le z < 1.0$ &   1.1   & 3.5     & 1.7     &  2.2 \\
&$1.0\le z < 1.1$ &   1.9   &         &         &  1.8 \\
&$1.1\le z < 1.2$ &   0.9   &         &         &\-0.7  \\
&$1.2\le z < 1.3$ &   4.3   &         &         &  2.7 \\
&$1.3\le z < 1.4$ & \-0.1   &         &         &  0.6 \\
\mr
$0.1\le z_{\mbox{\scriptsize nearby}}<0.2$ & 
 $0.2\le z < 0.3$ & \-0.8   & \-0.6   & 2.1     &   1.9 \\
&$0.3\le z < 0.4$ & \-0.2   & \-0.8   & 0.5     &   1.7 \\
&$0.4\le z < 0.5$ &   1.1   & \-0.5   & 1.9     &   3.5 \\
&$0.5\le z < 0.6$ &   1.1   &   0.7   & 3.3     &   3.8 \\
&$0.6\le z < 0.7$ &   1.1   &   0.0   & 0.7     &   2.4 \\
&$0.7\le z < 0.8$ & \-0.8   &   1.1   & 3.3     &   2.7 \\
&$0.8\le z < 0.9$ &   0.8   &   0.5   & 2.7     &   3.1 \\
&$0.9\le z < 1.0$ & \-0.8   &   0.9   & 0.9     &   1.9 \\
&$1.0\le z < 1.1$ &   1.1   &         &         &   1.6 \\
&$1.1\le z < 1.2$ & \-0.1   &         &         & \-1.0 \\
&$1.2\le z < 1.3$ &   2.6   &         &         &   2.3 \\
&$1.3\le z < 1.4$ & \-1.2   &         &         &   0.4 \\
\br
\end{tabular}
\end{indented}
\end{table}

\begin{table}
\begin{indented}
\lineup
\caption{\label{deltamutab} Evidence for acceleration
  $(\Delta\mu-\Delta\mu_{\mbox{\scriptsize nearby}})/\sigma$ for different data
  sets and fitting methods, where nearby SNe are defined as
  those SNe with redshifts fulfilling the given inequalities for
  $z_{\mbox{\scriptsize nearby}}$. Also given are the numbers of SNe
  in different redshift intervals.}
\item[] 
\begin{tabular}{@{}llllll}
\br
                                 & Gold    & ESSENCE & ESSENCE & Union \\
                                 &(MLCS2k2)&(MLCS2k2)& (SALT)  & (SALT)\\\mr
$0.0\le z_{\mbox{\scriptsize nearby}}<0.2$ & 4.3 & 5.2  & 5.6    & 7.2 \\
$0.0\le z_{\mbox{\scriptsize nearby}}<0.1$ & 4.4 & 5.7  & 5.7    & 5.9\\
$0.1\le z_{\mbox{\scriptsize nearby}}<0.2$ & 0.8 & 0.9  & 4.3    & 3.7 \\
\mr
\#SNe at $z<0.1$                  & \036   & \043    & \044   & \051 \\
\#SNe at $0.1\le z<0.2$           & \0\04  & \0\04   & \0\02  & \0\06 \\
\#SNe at $z>0.2$                  & 142    & 115     & 132    & 250 \\
\#SNe in total                    & 182    & 162     & 178    & 307 \\
\br
\end{tabular}
\end{indented}
\end{table}

Analysing the data sets using $0.0\le z_{\mbox{\scriptsize
    nearby}}<0.1$ and $0.1\le z_{\mbox{\scriptsize nearby}}<0.2$,
respectively, a drawback of this test becomes evident: It crucially
depends on the SNe in the first bin. (Note that this drawback
is also implicitly included in calibration-dependent tests as
nearby SNe are needed to determine $M$ and $H_0$.) Skipping the SNe
with a redshift smaller than 0.1, the evidence for acceleration
vanishes if the MLCS2k2 fitter is used, which could be partly due to
the fact that there are only four SNe classified as nearby. 
The reason why there is still some evidence if we use ESSENCE (SALT)
instead of ESSENCE (MLCS2k2) is not only due to the systematic effects
that occur when different light-curve fitters are used. The main reason
is that there are only two nearby SNe in ESSENCE (SALT), which
by chance have almost the same value of $\Delta\mu$ and thus a very
small variance. The Union set contains 6 nearby SNe with $z\ge 0.1$
and still states acceleration at $3.7\sigma$.

As large scale structures are observed up to $\sim400$Mpc (the Sloan
Great Wall \cite{Gott}) which corresponds to a redshift of about
$0.1$, it is questionable if the assumption of isotropy and
homogeneity is still justified for the analysis of SNe at lower
redshifts. Thus, one should prefer to make cosmological tests using
only SNe with higher redshifts and average over bin widths of at least
$\Delta z\ge0.1$. Unfortunately, this is not possible at
the moment as there are very few SNe at redshifts between 0.1 and 0.3
in all presently available data sets. As soon as SN data at
intermediate redshifts are published, repeating our analysis will show
if better statistics gives rise to at least some evidence of cosmic
acceleration and if there is still a significant difference in the
evidences obtained by using MLCS2k2 and SALT, respectively. 

\begin{figure}
\includegraphics[width=7.5cm]{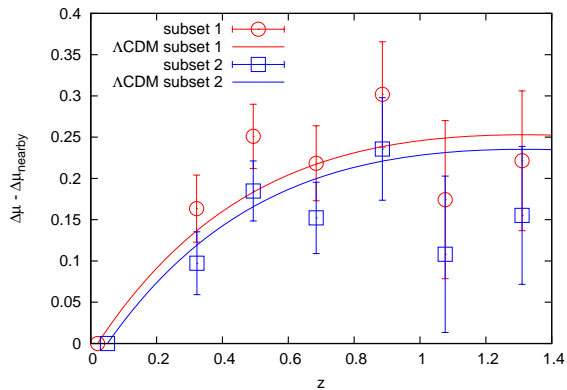}
\caption{$\Delta\mu-\Delta\mu_{\mbox{\scriptsize nearby}}$ for the
  Union set. The nearby SNe are those of subset 1 and 2, respectively,
  as given in table \ref{nearbysubset}.}
\label{split}
\end{figure}

In order to analyse more quantitatively how much the evidence changes
if the lowest redshifts are not considered, we split the SNe with
$z<0.1$ of each set into two subsets containing an equal number of
SNe. Subset 1 contains the SNe with the lowest redshifts, subset 2
those with the largest redshifts. $\Delta\mu_{\mbox{\scriptsize
    nearby}}$ is calculated using subset 1 and 2, respectively.
For the determination of $\Delta\mu$
we use all SNe with $z\ge0.2$. The result for the evidence for
acceleration is shown in table \ref{nearbysubset}. There is a tendency
of decreasing evidence when the lowest redshift SNe are
dismissed, which is not surprising. Only the Gold sample does not show
this trend. The amount by which the evidence is decreased is, however,
unexpected. This can be seen in figure \ref{split} for
the Union set: $\Delta\mu-\Delta\mu_{\mbox{\scriptsize
    nearby}}$ for a $\Lambda$CDM model ($\Omega_m=0.28$,
$\Omega_\Lambda=0.72$) using subset 2 drops only by 0.018 mag as
compared to the case when subset 1 is used, whereas the data points
drop by 0.066 mag. Thus, the change in the data points is much larger
than expected from $\Lambda$CDM. A model with a steeper increase of
$\Delta\mu(z)$ at small redshifts would be more consistent with these
data. An alternative explanation could be the so called Hubble bubble
or large void scenario, i.e. the local value of $H_0$ could be
different from the global one.

\begin{table}
\begin{indented}
\lineup
\caption{\label{nearbysubset} Evidence for acceleration
  $(\Delta\mu-\Delta\mu_{\mbox{\scriptsize nearby}})/\sigma$, where the SNe used
  to calculate $\Delta\mu$ have $z\ge 0.2$ and those to calculate
  $\Delta\mu_{\mbox{\scriptsize nearby}}$ have $z<0.1$. The nearby SNe are split
  into two subsets for each data set, each containing an equal number
  of SNe. Subset 1 contains the SNe with the smallest redshift, subset
  2 those with the largest redshifts. Also given is the weighted
  average redshift $\overline{z}_{\mbox{\scriptsize nearby}}$ and the
  number of SNe in each subset.}
\item[]
\begin{tabular}{@{}llllll}
\br
                 &  & Gold    & ESSENCE & ESSENCE & Union \\
                 &  &(MLCS2k2)&(MLCS2k2)& (SALT)  & (SALT)\\\mr
subset 1 & evidence & 3.1     & 6.6     & 5.6     & 6.3 \\
         & $\overline{z}_{\mbox{\scriptsize nearby}}$ 
                    & 0.030   & 0.021   & 0.021   & 0.021 \\
subset 2 & evidence & 3.4     & 3.2     & 3.9     & 4.8 \\
         & $\overline{z}_{\mbox{\scriptsize nearby}}$ 
                    & 0.056   & 0.046   & 0.050   & 0.051 \\\mr
\# SNe &            & 18      & 21      & 22      & 25 \\
\br
\end{tabular}
\end{indented}
\end{table}

\section{Open and closed universes}\label{open-closed}

We now give up the assumption of spatial flatness. Then the distance
modulus of a universe with deceleration parameter $q=0$ is given by
\begin{equation}
\mu_{q=0}(z) = 5\log\left[ \frac{1+z}{H_0\sqrt{|\Omega_k|}}\;
  \mathcal{S}\left[\sqrt{|\Omega_k|}\ln(1+z)\right] \right] + 25\;,
\end{equation}
where $\mathcal{S}(x)=\sin (x)$ for a closed and $\mathcal{S}(x)=\sinh
(x)$ for an open universe. Using this, $\Delta\mu_i = \mu_i -
\mu_{q=0}(z_i)$ and subsequently $\Delta\mu -
\Delta\mu_{\mbox{\scriptsize nearby}}$ can be easily
calculated. Defining the nearby SNe as those with $z<0.2$, the result
for different values of the spatial curvature $\Omega_k$
is given in table \ref{open}. 

In a closed universe, $\mu_{q=0}(z)$ decreases with increasing
curvature, i.e. $\Delta\mu - \Delta\mu_{\mbox{\scriptsize nearby}}$
(and thus the evidence for accelerated expansion) increases. As we are
interested in the lower limit of the evidence in a closed universe, we
need to consider the smallest possible curvature, i.e. $\Omega_k\to 0$.
Therefore, the results of the test are equal to those of a spatially
flat universe. In an open universe the
evidence decreases with increasing curvature. Thus, we need the
largest possible value, $\Omega_k=1$, to determine the lower limit to the
evidence. Note that $\Omega_k=1$ is the largest possible curvature
only if the rule $\sum_i\Omega_i=1$ holds, where
$i=m,\Lambda,k,\ldots$. This rule is, however, only valid for general
relativity and can be different in a modified gravity scenario.
These weakest evidences for a flat/closed and an open
universe are highlighted in table \ref{open}.

\begin{table}
\begin{indented}
\lineup
\caption{\label{open} Evidence for acceleration assuming different
  values of the spatial curvature $\Omega_k$. The lower limits to the
  evidence for a flat/closed and an open universe are
  highlighted. Nearby SNe are defined as those with redshift $z<0.2$.}
\item[]
\begin{tabular}{@{}llllll}
\br
                &           & Gold    & ESSENCE & ESSENCE & Union \\
                &$\Omega_k$ &(MLCS2k2)&(MLCS2k2)& (SALT)  & (SALT)\\\mr
closed universe &\-1.0      & 6.7     & 7.0     & 7.5     & 10.2 \\
                &\-0.8      & 6.3     & 6.6     & 7.1     & \09.6 \\
                &\-0.6      & 5.8     & 6.3     & 6.8     & \09.0 \\
                &\-0.4      & 5.3     & 5.9     & 6.4     & \08.4 \\
                &\-0.2      & 4.8     & 5.6     & 6.0     & \07.8 \\
flat universe   &  0.0      &{\bf 4.3}&{\bf 5.2}&{\bf 5.6}&{\bf\07.2} \\
open universe   &  0.2      & 3.8     & 4.9     & 5.3     & \06.6 \\
                &  0.4      & 3.3     & 4.5     & 4.9     & \06.0 \\
                &  0.6      & 2.8     & 4.1     & 4.5     & \05.4 \\
                &  0.8      & 2.3     & 3.7     & 4.1     & \04.8 \\
                &  1.0      &{\bf 1.8}&{\bf 3.4}&{\bf 3.8}&{\bf\04.2} \\
\br
\end{tabular}
\end{indented}
\end{table}

\section{Conclusion}
We modified our model-independent test of accelerated expansion
presented in \cite{Seikel} in such a way that systematics due to
calibrations are avoided. This could be achieved by considering the SN
data relative to the data in a bin of nearby SNe, where the
most ``natural'' choice is to define the nearby SNe as those with
redshift $z<0.2$. Compared to
the previous test, the evidence for acceleration is weaker, but all
systematic effects are reduced. However, the evidence obtained from the
ESSENCE set is still larger than that obtained from the Gold sample,
and the SALT fitter gives a stronger evidence than MLCS2k2. Being
conservative, one should take the lowest evidence obtained by using
different data sets and light-curve fitters, i.e. 4.3$\sigma$ in the
case of a spatially flat universe using the Gold sample fitted with
MLCS2k2. The Union set is however not an independent data set, but
contains SNe from Gold, ESSENCE and other sets. Thus, the Union set
seems to be a good choice in order to determine the evidence for
accelerated expansion. Unfortunately, the published data of this set
have only been fitted with SALT. As for the ESSENCE set MLCS2k2
gives a weaker evidence than SALT, it is quite probable that this is
also the case for the Union set. Then MLCS2k2 would give a
more conservative value than the obtained 7.2$\sigma$ evidence. 

By changing the set of nearby SNe, it becomes obvious that the test
crucially depends on SNe with redshift $z\lesssim 0.1$. Skipping those
data leads to vanishing evidence when the MLCS2k2 light-curve fitter
is used. As a redshift of 0.1 corresponds approximately to a distance
of 400Mpc, which is the size of the largest observed structures in the
universe, the assumption of homogeneity and isotropy might not be
justified for such low redshifts. For the Union set (based on 6 SNe
between 0.1 and 0.2) still $3.7\sigma$ evidence is found. It remains
to be seen if this is confirmed by larger data sets in the future.

We conclude that the largest publicly available supernova data sets
show statistically significant evidence for cosmic acceleration 
if statistical isotropy and homogeneity are assumed at all redshifts. 
A violation of isotropy and/or homogeneity, e.g.~by a local void,
remains a viable alternative interpretation of supernova data (see
e.g.~\cite{CelerierRev}). 

How to proceed? First of all we need more data at $0.1<z<0.3$. This
would allow to anchor our test with supernovae at intermediate
redshifts, which should not be affected by the details of the local
environment. Thus such a test could detect cosmic acceleration in a 
way that relies only on the assumptions of isotropy and homogeneity on
scales at which they can be established by observations directly. 
Secondly, in \cite{SW} a significant anisotropy of Hubble diagrams on
opposite hemispheres of the sky has been found. The direction of maximal
asymmetry was found to be close to the pole of the equatorial coordinate
system, which points towards systematic errors in existing data
sets. Thus with improved data sets, especially better control on
systematics at all levels, one could try to use SN Ia at $z<0.1$ to 
establish isotropy and homogeneity at even smaller redshifts. 
In both cases, ideally full sky surveys for nearby supernovae are
required, while pencil beam like approaches would make it more
difficult to disentangle a local void from dark energy.  

\ack
We thank Marek Kowalski and Adam Riess for discussions and comments.
This work is supported by the DFG under grant GRK 881.

\end{document}